\documentclass[11pt,a4paper]{amsart}
\usepackage[utf8]{inputenc}
\usepackage{amsmath}
\usepackage{amsfonts}
\usepackage{amssymb}
\usepackage{graphicx}
\usepackage{booktabs}

\usepackage[
left=1.8cm, 
right=1.8cm, 
top=1.8cm, 
bottom=1.8cm 
]{geometry}

\numberwithin{equation}{section}  
\newtheorem{defn}{Definition}[section]
\newtheorem{exm}[defn]{Example}
\newtheorem{remark}[defn]{Remark}
\newtheorem{thm}[defn]{Theorem}

\title
{A Note on Optimal Liquidation with Linear Price Impact}

\author{Yan Dolinsky \and Doron Greenstein}
\address{Hebrew University, Department of Statistics, Mount Scopus, Jerusalem, Israel.}
\begin{document}
\vspace{-0.5cm}

\begin{abstract}
In this note we consider the maximization of the expected terminal wealth for the setup of quadratic transaction costs.
First, we provide a very simple probabilistic solution to the problem. Although the problem was largely studied, as far as we know
up to date this simple and probabilistic form of the solution has not appeared in the literature.
 Next, we apply the general result for the numerical study of
the case where the risky asset is given by a fractional Brownian Motion and the information flow of the
investor can be diversified.
\\
\\
\textbf{Keywords: }{Linear Price Impact, Optimal Liquidation, Fractional Brownian Motion}
\end{abstract}

\begin{minipage}[t]{\textwidth}
\vspace*{-4\baselineskip}
\maketitle
\end{minipage}

\section{Preliminaries and the General Result}
\noindent
Consider a model
with one risky asset which we denote by $S=(S_t)_{0\leq t\leq T}$ , where $T<\infty$ is the
time horizon. We assume that the investor has a bank account that, for simplicity,
bears no interest. The risky asset $S$ is
 RCLL (right continuous with left limits) and adapted process defined on a filtered
probability space
$(\Omega, \mathcal{F},(\mathcal F_t)_{0\leq t\leq T}, \mathbb P)$.
The filtration $(\mathcal F_t)_{0\leq t\leq T}$ satisfies the
usual assumptions (right continuity and completeness). 
Let us emphasize that we do not assume that the $\sigma$-algebra $\mathcal F_0$ 
is the trivial $\sigma$-algebra. 

In financial markets, trading moves prices against the trader:
buying faster increases execution prices, and selling faster decreases them. This aspect
of liquidity, known as market depth (see \cite{B})
 or price-impact, has received large
attention in optimal liquidation problems,
see, for instance,
\cite{AlmgrenChriss:01,GS:11,BV:2019,FSU:19}
and the references therein.

Following \cite{AlmgrenChriss:01}, we model the investor’s market impact in a
temporary linear form and thus, when at time $t$ the investor turns over her position $\Phi_t$ at the rate $\phi_t=\dot{\Phi}_t$
the execution price is $S_t+\frac{\Lambda}{2}\phi_t$ for some constant $\Lambda>0$. In our setup the investor has to liquidate
his position, namely $\Phi_T=\Phi_0+\int_{0}^T\phi_t dt=0$.
For a given initial number (deterministic) of shares $\Phi_0$, denote by $\mathcal A_{\Phi_0}$ the set of all progressively measurable processes 
$\phi=(\phi_t)_{0\leq t\leq T}$
which satisfy 
$\int_{0}^T \phi^2_t  dt<\infty$ and $\Phi_0+\int_{0}^T \phi_t dt=0.$
As usual, all the equalities and the inequalities are understood in the almost surely sense.

The profits and
losses from trading are given by
\begin{equation}\label{2.1}
V^{\Phi_0,\phi}_T:=-\Phi_0 S_0-\int_{0}^T \phi_t S_t dt-\frac{\Lambda}{2}\int_{0}^T \phi^2_t dt.
\end{equation}
Observe that for $\phi\in\mathcal A_{\Phi_0}$ the right hand side of (\ref{2.1}) is well defined 
if $\int_{0}^T S^2_t dt<\infty$. This inequality follows from the integrability condition given by (\ref{2.0}). 
In particular, we do not assume that $S$ is a semi--martingale.

Let us explain in more detail formula (\ref{2.1}). At time $0$ the investor has $\Phi_0$ stocks and
the sum $-\Phi_0 S_0$ on her savings account. At time $t\in [0,T)$
the investor buys $\phi_t dt$, an infinitesimal number of stocks or more intuitively
sell $-\phi_t dt$ number of shares and so the (infinitesimal) change in the savings account
is given by $-\phi_t \left(S_t+\frac{\Lambda}{2}\phi_t\right)dt$. Since we liquidate
the portfolio at the maturity date, the terminal portfolio value is equal to
the terminal amount on the savings account and given by
$-\Phi_0 S_0- \int_{0}^T \phi_t \left(S_t+\frac{\Lambda}{2}\phi_t\right)dt$. We arrive at the right-hand side of (\ref{2.1}).
For the case where $S$ is a semi--martingale, by applying the integration by parts formula
$\int_{0}^T \Phi_t dS_t=\Phi_T S_T-\Phi_0 S_0-\int_{0}^T S_td\Phi_t$
and using the fact that $\Phi_T=0$ (liquidation)
 we get that the right-hand side of (\ref{2.1}) is equal to
$\int_{0}^T \Phi_t dS_t-\frac{\Lambda}{2} \int_{0}^T \phi^2_t dt$.

We are interested in the following optimal liquidation problem 
\begin{equation}\label{2.1+}
\mbox{Maximize} \  \mathbb E\left[V^{\Phi_0,\phi}_T\right] \ \ \mbox{over} \ \ \phi\in\mathcal A_{\Phi_0}
\end{equation}
where $\mathbb E$ denotes the expectation with respect to $\mathbb P$.

The following theorem provides a complete probabilistic solution to the optimization problem 
(\ref{2.1+}). 
\begin{thm}\label{thm2.1}
Assume that
\begin{equation}\label{2.0}
\mathbb E\left[\int_{0}^T S^2_t dt\right]<\infty.
\end{equation}
Introduce the martingale 
\begin{equation}\label{mart}
M_t:=\mathbb E\left[\int_{0}^T S_u du\left |\right. \mathcal F_t\right] \ \ t\in [0,T].
\end{equation}
The unique ($dt\otimes\mathbb P$ a.s) solution to the optimization problem (\ref{2.1+}) is given by 
\begin{equation}\label{port}
\hat\phi_t:=-\frac{\Phi_0}{T}+\frac{M_0}{T\Lambda}+
\frac{1}{\Lambda}\left(\int_{0}^t \frac{dM_u}{T-u}-S_t\right), \ \ t\in [0,T)
\end{equation}
and the corresponding value is equal to 
\begin{eqnarray}\label{val}
&\max_{\phi\in\mathcal A_{\Phi_0}}\mathbb E\left[V^{\Phi_0,\phi}_T\right]=\mathbb E\left[V^{\Phi_0,\hat\phi}_T\right] \nonumber\\
&=-\frac{\Phi^2_0 \Lambda}{2 T}+\Phi_0\mathbb E\left[\frac{M_0}{T}-S_0\right]+  
\frac{1}{2\Lambda} \mathbb E\left[\int_{0}^T\left(S_t-\frac{M_0}{T}-\int_{0}^t \frac{dM_u}{T-u}\right)^2 dt
\right].
\end{eqnarray}
\end{thm}
A slightly more general form of the linear-quadratic optimization problem (\ref{2.1+}) has been considered in
\cite{BV}, however for the relatively simple setup of 
optimal liquidation Theorem \ref{thm2.1} provides a much simpler solution than \cite{BV}.
As far as we know, up to date this simple and probabilistic form of the solution has not appeared in
the literature. 

Before, we prove Theorem \ref{thm2.1} let us briefly collect some observations from this result.
First, let us notice that it is sufficient to define the optimal portfolio on the
half-open interval $[0,T)$  (as we do in (\ref{port}).
 We can just set $\phi_T:=0$.

Next, observe that the optimal value given by the right hand side of (\ref{val}) can be decomposed into three terms, the first 
$-\frac{\Phi^2_0 \Lambda}{2 T}$ does not depend on the risky asset, the second term is a product of the initial number of shares 
$\Phi_0$ and the term $\mathbb E\left[\frac{M_0}{T}-S_0\right]$ which can be interpreted
as the average drift of the risky asset $S$ (recall that we do not assume that $S$ is a semi-martingale). The last term 
$\frac{1}{2\Lambda} \mathbb E\left[\int_{0}^T\left(S_t-\frac{M_0}{T}-\int_{0}^t \frac{dM_u}{T-u}\right)^2 dt\right]$
is a product of the market depth $\frac{1}{2\Lambda}$ and the distance of the risky asset $S$ from a martingale. 
In particular if $S$ is a martingale then the last term is zero. Indeed, if $S$ is a martingale then (\ref{mart}) implies
$M_t=\int_{0}^t S_u du+(T-t)S_t$, $t\in [0,T]$. From 
the (stochastic) Leibniz rule we get
$dM_t=S_t dt+(T-t) dS_t- S_t dt=(T-t)dS_t$. This together with the equality 
$\frac{M_0}{T}=S_0$ gives $S_t=\frac{M_0}{T}+\int_{0}^t \frac{dM_u}{T-u}$ for all $t$. 
 
Next, we prove Theorem \ref{thm2.1}.
\begin{proof}
The proof will be done in three steps. \\
\textbf{Step I:}  
Introduce the process 
$N_t:=\int_{0}^t  \frac{dM_u}{T-u}$, $t\in [0,T).$
In this step we show that 
\begin{equation}\label{2.10}
\mathbb E\left[\int_{0}^T S_t N_t dt\right]=\mathbb E\left[\int_{0}^T N^2_t dt\right]\leq
\mathbb E\left[\int_{0}^T S^2_t dt\right].
\end{equation}
Fix $n\in\mathbb N$ and define the process $N^n=(N^n_t)_{0\leq t\leq T}$ by 
$N^n_t:=N_{t\wedge (T-1/n)}$, $t\in [0,T]$.
From (\ref{2.0}) it follows that $M$ and $N^n$ are square integrable martingales. 

Next, for any square integrable martingales $X,Y$ we denote by $[X]$ the quadratic variation of $X$ and by $[X,Y]$ the covariation 
of $X$ and $Y$. 
Also, denote by $\mathbb I_{\cdot}$ the indicator function. 

Observe that,
\begin{eqnarray*}\label{2.11}
&\mathbb E\left[\int_{0}^T S_t N^n_{t} dt\right]=\mathbb E\left[N^n_T\int_{0}^T S_t dt \right]=\mathbb E\left[M_T N^n_T\right]\\
&=\mathbb E\left[[M,N^n]_T\right]=\mathbb E\left[\int_{0}^T\frac{\mathbb I_{s<T-1/n}}{T-s}d[M]_s\right]\nonumber\\
&=\mathbb E\left[\int_{0}^T\int_{s}^T\frac{\mathbb I_{s<T-1/n}}{(T-s)^2}dtd[M]_s\right]\\
&=\mathbb E\left[\int_{0}^T\int_{0}^t\frac{\mathbb I_{s<T-1/n}}{(T-s)^2}d[M]_sdt\right]
=
\mathbb E\left[\int_{0}^T |N^n_t|^2 dt\right].
\end{eqnarray*}
Indeed, the first equality follows from the fact that $N^n$ is a square integrable martingale. The second equality is due to (\ref{mart}).
 The third equality follows from Theorem 6.28 in \cite{H} (we note that $N^n_0=0$). 
The fourth equality follows from Theorem 9.15 in \cite{H} where the integral with respect to $d[M]$ is the (pathwise) 
Stieltjes integral with respect to the non decreasing process $[M]$. The fifth equality is obvious. 
 The sixth equality is due to the Fubini theorem. 
Finally, the last equality is due to the (generalized) It\^{o} Isometry (see Chapter IX in \cite{H})
 which says that for any bounded and predictable process $\mathcal H$ and a square integrable martingale $X$ we have
 $\mathbb E\left[\left(\int_{0}^T H_t dX_t\right)^2\right]=\mathbb E\left[\int_{0}^T H^2_t d[X]_t\right].$
 
We conclude
\begin{equation}\label{2.11}
\mathbb E\left[\int_{0}^T S_t N^n_{t} dt\right]=
\mathbb E\left[\int_{0}^T |N^n_t|^2 dt\right].
\end{equation}
Hence,
\begin{equation}\label{2.12}
0\leq \mathbb E\left[\int_{0}^T|S_t- N^n_t|^2 dt\right]=
\mathbb E\left[\int_{0}^T S^2_t dt\right]- \mathbb E\left[\int_{0}^T |N^n_t|^2 dt\right].
\end{equation}
From (\ref{2.0}) and (\ref{2.11})--(\ref{2.12}) we obtain 
\begin{eqnarray*}
&\mathbb E\left[\int_{0}^T S_t N_t dt\right]=\lim_{n\rightarrow\infty}
\mathbb E\left[\int_{0}^T S_t N^n_t dt\right]
\\
&=\lim_{n\rightarrow\infty}
\mathbb E\left[\int_{0}^T |N^n_t|^2 dt\right]\\
&=\mathbb E\left[\int_{0}^T N^2_t dt\right]\leq
\mathbb E\left[\int_{0}^T S^2_t dt\right]
\end{eqnarray*}
and (\ref{2.10}) follows. 
\\${}$\\
  \textbf{Step II:}
  Let $\phi\in\mathcal A_{\Phi_0}$. In this step we prove that $\mathbb E\left[V^{\Phi_0,\phi}_T\right]$ is not bigger than the right hand side 
  of (\ref{val}). Without loss of generality we assume that
   $\mathbb E\left[V^{\Phi_0,\phi}_T\right]>-\infty$. 
   
   From (\ref{2.1}) and the
Cauchy–Schwarz inequality
it follows
that
$$
\sqrt{\int_{0}^T S^2_t dt}\sqrt{\int_{0}^T \phi^2_t dt}  -\frac{\Lambda}{2}\int_{0}^T \phi^2_t dt\geq
V^{\Phi_0,\phi}_T+\Phi_0S_0.$$
Thus, 
$$\frac{\Lambda}{2}\left(\sqrt{\int_{0}^T \phi^2_t dt}-\frac{1}{\Lambda}\sqrt{\int_{0}^T S^2_t dt}\right)^2
\leq \frac{1}{2\Lambda} \int_{0}^T S^2_t dt-V^{\Phi_0,\phi}_T-\Phi_0S_0.
$$
This together with the integrability condition (\ref{2.0}) and the inequality 
$\mathbb E\left[V^{\Phi_0,\phi}_T\right]>-\infty$ gives that 
$\sqrt{\int_{0}^T \phi^2_t dt}-\frac{1}{\Lambda}\sqrt{\int_{0}^T S^2_t dt}\in L^2(\mathbb P)$. Clearly, \
(due to (\ref{2.0}))  
$\sqrt{\int_{0}^T S^2_t dt}\in L^2(\mathbb P)$, and so we conclude that 
$\sqrt{\int_{0}^T \phi^2_t dt}\in L^2(\mathbb P)$, i.e. 
$
\mathbb E\left[\int_{0}^T\phi^2_t dt\right]<\infty. 
$

Next, set 
$Z:=-\frac{\Phi_0\Lambda }{T}+\frac{M_0 }{T}$ and
choose $n\in\mathbb N$. 
From the estimate $
\mathbb E\left[\int_{0}^T\phi^2_t dt\right]<\infty
$ and the fact that $N^n$ is square integrable martingale we obtain 
$$\mathbb E\left[\int_{0}^T \phi_t N^n_t dt\right]=\mathbb E\left[N^n_T\int_{0}^T \phi_t dt \right]=-\Phi_0
\mathbb E\left[ N^n_T\right]=0.$$
 This together with 
(\ref{2.1}) and the simple inequality
$x y-\frac{\Lambda}{2}x^2\leq \frac{y^2}{2\Lambda}$, $x,y\in\mathbb R$ yields
\begin{eqnarray*}
&\mathbb E\left[V^{\Phi_0,\phi}_T\right]=\mathbb E\left[-\Phi_0 (S_0-Z)-\int_{0}^T \phi_t (S_t-Z-N^n_t)  dt-\frac{\Lambda}{2}\int_{0}^T \phi^2_t dt\right]\nonumber\\
&\leq \mathbb E\left[-\Phi_0 (S_0-Z)+\frac{1}{2\Lambda}\int_{0}^T |S_t-Z-N^n_t|^2 dt\right].
\end{eqnarray*}
By taking $n\rightarrow\infty$ in the above inequality and applying (\ref{2.10})
we obtain 
\begin{equation}\label{2.20}
\mathbb E\left[V^{\Phi_0,\phi}_T\right]\leq -\frac{\Phi^2_0 \Lambda}{2 T}+\Phi_0\mathbb E\left[\frac{M_0}{T}-S_0\right]
+  
\frac{1}{2\Lambda} \mathbb E\left[\int_{0}^T\left(S_t-\frac{M_0}{T}- N_t\right)^2 dt
\right]
\end{equation}
as required.
\\${}$\\
\textbf{Step III:}
In this step we complete the proof.
Consider the trading strategy given by (\ref{port}).
From the Fubini theorem it follows that 
$$\int_{0}^T \hat\phi_t dt=-\Phi_0+\frac{1}{\Lambda}\left(M_0
+M_T-M_0-\int_{0}^T S_t dt\right)=-\Phi_0.$$
Moreover, from (\ref{2.10}) it follows that 
$\mathbb E\left[\int_{0}^T\hat\phi^2_t dt\right]<\infty$.
Thus, $\hat\phi\in\mathcal A_{\Phi_0}$.

Next, choose $n\in\mathbb N$. 
By using the same arguments as in Step II we get
$\mathbb E\left[\int_{0}^T \hat\phi_t N^n_t dt\right]=0.$
Observe that for 
$t\leq T-1/n$ we have $\hat\phi_t=\frac{Z+N^n_t-S_t}{\Lambda}$, where (recall) $Z=-\frac{\Phi_0\Lambda }{T}+\frac{M_0 }{T}$.
Hence,
\begin{eqnarray*}
&\mathbb E\left[V^{\Phi_0,\hat\phi}_T\right]=\mathbb E\left[-\Phi_0 (S_0-Z)-\int_{0}^T \hat\phi_t (S_t-Z-N^n_t)  dt-\frac{\Lambda}{2}\int_{0}^T \hat\phi^2_t dt\right]\nonumber\\
&= \mathbb E\left[-\Phi_0 (S_0-Z)+\frac{1}{2\Lambda}\int_{0}^{T-1/n} |S_t-Z-N_t|^2 dt\right]\\
&-\mathbb E\left[\int_{T-1/n}^T \hat\phi_t (S_t-Z-N^n_t)  dt+\frac{\Lambda}{2}\int_{T-1/n}^T \hat\phi^2_t dt\right].
\end{eqnarray*}
By taking $n\rightarrow\infty$ in the above equality and applying (\ref{2.10})
we obtain (notice that $\mathbb E\left[\int_{0}^T\hat\phi^2_t dt\right]<\infty$)
\begin{eqnarray}\label{2.21}
&\mathbb E\left[V^{\Phi_0,\hat\phi}_T\right]=\mathbb E\left[-\Phi_0 (S_0-Z)+\frac{1}{2\Lambda}\int_{0}^{T} |S_t-Z-N_t|^2 dt\right]\nonumber\\
&= -\frac{\Phi^2_0 \Lambda}{2 T}+\Phi_0\mathbb E\left[\frac{M_0}{T}-S_0\right]
+  
\frac{1}{2\Lambda} \mathbb E\left[\int_{0}^T\left(S_t-\frac{M_0}{T}- N_t\right)^2 dt
\right].
\end{eqnarray}
By combining (\ref{2.20})--(\ref{2.21}) we conclude (\ref{val}).

Finally, the 
uniqueness of the optimal trading strategy follows from the strict convexity of the map $\phi\rightarrow V^{\Phi_0,\phi}_T$.
\end{proof}

We end this section with the following example. 
\begin{exm}
Assume that $S$ is a square integrable martingale with respect to the filtration 
 $(\mathcal F_t)_{0\leq t\leq T}$. By applying the 
 same arguments as in the paragraph before the proof of Theorem \ref{thm2.1}, we obtain 
 that $S_t=\frac{M_0}{T}+\int_{0}^t \frac{dM_u}{T-u}$, $t\in [0,T]$.
 This together with (\ref{port}) gives that the optimal strategy is purely deterministic and equals to
$\hat\phi_t\equiv -\frac{\Phi_0}{T}$. Namely, we liquidate our initial position $\Phi_0$ at a constant rate. 
From (\ref{val}) we obtain that the corresponding value is equal to $-\frac{\Phi^2_0\Lambda}{2 T}$. 
Since $\hat\phi$ is deterministic, then in the case of partial information, i.e. where the investor's filtration is smaller than $(\mathcal F_t)_{0\leq t\leq T}$,
the solution to the optimization problem (\ref{2.1+}) will be the same. 

A more interesting case is where the filtration is larger than 
 $(\mathcal F_t)_{0\leq t\leq T}$. More precisely, fix $\Delta\in (0,T]$ and consider the case where the investor
can peek $\Delta$ time units into the future, and so her information
flow is given by the filtration
$(\mathcal F_{t+\Delta})_{t\geq 0}$. 

From (\ref{mart}) we obtain that 
$$
M_t=\int_{0}^{{(t+\Delta)}\wedge T}S_u du+(T-t-\Delta)^{+}S_{(t+\Delta)\wedge T}, \ \ t\in [0,T].$$
Thus, 
$M_0=\int_{0}^{\Delta} S_u du+(T-\Delta) S_{\Delta}$
and from the 
 Leibniz rule we get 
 \begin{eqnarray*}
 &dM_t=\mathbb I_{t<T-\Delta} \left(S_{t+\Delta}dt+\left(T-t-\Delta\right)dS_{t+\Delta}-S_{t+\Delta}dt\right)\\
 &=\mathbb I_{t<T-\Delta} \left(T-t-\Delta\right)dS_{t+\Delta}, \ \ t\in [0,T].
 \end{eqnarray*}
 Hence,
 \begin{eqnarray*}
 &\frac{M_0}{T}+\int_{0}^t\frac{dM_u}{T-u}-S_t\\
 &=\frac{1}{T}\left(\int_{0}^{\Delta} S_u du+(T-\Delta) S_{\Delta}\right)+\int_{\Delta}^{(t+\Delta)\wedge T}\frac{T-u}{T+\Delta-u}dS_u-S_t\\
 &=\frac{\int_{0}^{\Delta} (S_u-S_{\Delta}) du}{T}
 +S_{(t+\Delta)\wedge T}-S_t-\Delta
 \int_{\Delta}^{(t+\Delta)\wedge T}\frac{ dS_u}{T+\Delta-u}, \ \ t\in [0,T].
 \end{eqnarray*}
 
 This together with (\ref{port})-(\ref{val}) yields that the optimal strategy is given by
 $$\hat\phi_t=-\frac{\Phi_0}{T}+\frac{\int_{0}^{\Delta} (S_u-S_{\Delta}) du}{T\Lambda}
 +\frac{S_{(t+\Delta)\wedge T}-S_t}{\Lambda}-\frac{\Delta}{\Lambda}
 \int_{\Delta}^{(t+\Delta)\wedge T}\frac{dS_u}{T+\Delta-u}$$
and the corresponding value (notice that $\mathbb E[M_0]=S_0 T$) is
 equal to 
$$\mathbb E\left[V^{\Phi_0,\hat\phi}_T\right]=-\frac{\Phi^2_0\Lambda}{2T}+\frac{I}{2\Lambda}$$
where 
$$I:=\mathbb E\left[\int_{0}^T\left(\frac{\int_{0}^{\Delta} (S_u-S_{\Delta}) du}{T}
 +S_{(t+\Delta)\wedge T}-S_t-\Delta
 \int_{\Delta}^{(t+\Delta)\wedge T}\frac{dS_u}{T+\Delta-u}\right)^2 dt\right]$$
 can be viewed as the premium of being able to peek ahead by $\Delta$ units of time.
 \end{exm}
\section{The Case of Fractional Brownian Motion}
Fractional Brownian motion 
$B^H=(B^H_t)_{t=0}^{\infty}$
with Hurst parameter $H\in (0,1)$, is
a continuous, zero-mean Gaussian process such that
\begin{equation*}
cov\left(B^H_t,B^H_u\right)=\frac{t^{2H}+u^{2H}-|t-u|^{2H}}{2}, \ \ t,u\geq 0.
\end{equation*}
The process $B^H$ is self similar $B^H_{at}\sim a^H B^H_t$ and  
have stationary increments. Moreover, the successive 
increments of $B^H$ are positively correlated for $H>1/2$, negatively correlated for $H<1/2$, while $H = 1/2$ 
recovers the usual Brownian motion with independent increments.

Fractional Brownian motion which displays the long-range dependence observed in empirical date
(see \cite{CKW,M,WTT} and the references therein)
is not a semi-martingale when 
$H\neq \frac{1}{2}$ and so, in the frictionless case it leads to arbitrage opportunities (see, for instance, 
\cite{R,C}).
In the presence of market price impact arbitrage opportunities disappear 
and the expected profits are finite 
(see \cite{G1,G2}). 
In \cite{G2} the authors studied
 the asymptotic behaviour (as the maturity date goes to infinity)
of the optimal liquidation problem with temporary price impact,
for the case where the risky asset is given by
 a fractional Brownian motion. It is also important to mention the recent paper \cite{MISH} which is closely related. 

In this section, for the financial model where the risky asset is given by a fractional Brownian motion, we study the dependence
of the optimal liquidation problem
as a function of the investor's information. We deal with three types of investors.
 The first one, is the "usual" investor with information flow which is given by the filtration generated by the risky asset.
 The second type is an investor which receives the information with a delay. The last type is a "frontrunner"
 which is able to peek some time units into the future.
Of course the "frontrunner" cannot freely take advantage of her extra knowledge due
to the linear price impact which leads to quadratic transaction costs. For the above three cases
we solve the corresponding optimal liquidation problem and derive numerical results for the value (see Figure 1) and
for the optimal strategy (see Figure 2). 
 
Let $H\in (0,1)$ and consider the optimization problem (\ref{2.1+}) for the case where the risky asset is of the form 
$S_t=S_0+\sigma B^H_t+\mu t$ where $\sigma>0$ and 
$\mu\in\mathbb R$ are constants. From Theorem \ref{thm2.1} and the discussion afterwards it follows 
that (for simplicity) we can take 
$\mu=S_0=0$ and $\sigma=\Lambda=1$. Thus, $S=B^H$ for some $H\in (0,1)$
and $\Lambda=1$. 

For $H\in (0,1)$ introduce the Volterra kernel
\begin{eqnarray*}
&Z_H(t,s)=c_H
\left(\left(\frac{t}{s}\right)^{H-\frac{1}{2}}(t-s)^{H-\frac{1}{2}}\right.\\
&-\left.\left(H-\frac{1}{2}\right)s^{\frac{1}{2}-H}\int_{s}^tu^{H-\frac{3}{2}}(u-s)^{H-\frac{1}{2}}du\right), \ \ 0<s<t
\end{eqnarray*}
where 
$c_H:=\left(\frac{2 H\Gamma\left(\frac{3}{2}-H\right)}{\Gamma\left(H+\frac{1}{2}\right)\Gamma(2-2H)}\right)^{1/2}.$
Then, taking an ordinary Brownian motion $W=(W_t)_{t=0}^{\infty}$ the formula
\begin{equation}\label{3.1}
B^H_t=\int_{0}^t Z_H(t,s) dW_s, \ \ t\geq 0.
\end{equation}
defines a fractional Brownian motion 
with Hurst parameter $H$, which generates the same filtration as $W$ (see \cite{Val}).
Moreover, given $B^H$, the Wiener process $W$ can be recovered by 
the relations 
$$W_t:=\frac{2H}{c_H}\int_{0}^t s^{H-\frac{1}{2}}d\mathcal M_s, \ \ t\geq 0$$ where 
$$\mathcal M_t:=\frac{1}{2H \Gamma\left(\frac{3}{2}-H\right)\Gamma\left(H+\frac{1}{2}\right)}
\int_{0}^t s^{\frac{1}{2}-H}(t-s)^{\frac{1}{2}-H} dB^H_s, \ \ t\geq 0.$$
Denote by $(\mathcal F^W_t)_{t\geq 0}$ the augmented filtration which is generated by $W$. 

\subsection{Standard Information}
Consider the case where the filtration $(\mathcal F_t)_{0\leq t\leq T}$ (which represent the investor's flow of information)
is equal to $(\mathcal F^W_t)_{0\leq t\leq T}$.
From the Fubini theorem and (\ref{3.1}) it follows that the martingale defined in (\ref{mart}) is equal to 
$$M^H_t=\int_{0}^t  \left(\int_{s}^T Z_H(u,s)du\right)dW_s, \ \ t\in [0,T].$$ 
Hence,  
(\ref{port}) and (\ref{3.1}) yield that the optimal strategy is given by
\begin{equation*}\label{3.2}
\hat\phi^H_t:=\int_{0}^t \left(\frac{\left(\int_{s}^T Z_H(u,s)du\right)}{T-s}-Z_H(t,s)\right)dW_s, \ \ t\in [0,T].
\end{equation*}
From the It\^{o} Isometry and (\ref{val}) we obtain that the corresponding value is given by 
\begin{eqnarray*}\label{3.3}
&\mathbb E\left[V^{0,\hat\phi^H}_T\right]=
\int_{0}^T \int_{0}^t  Z^2_H(t,s) ds dt -\int_{0}^T  \frac{\left(\int_{s}^T Z_H(u,s)du\right)^2}{T-s} ds\\
&=\frac{T^{2H+1}}{2H+1}-\int_{0}^T  \frac{\left(\int_{s}^T Z_H(u,s)du\right)^2}{T-s} ds.
\end{eqnarray*}

\subsection{Delayed Information}
 We fix a positive number $\Delta\in (0,T]$ and consider a situation where
the risky asset $S$ is observed with a delay $\Delta>0$. Namely, the filtration is
$\mathcal F_t=\mathcal F^W_{(t-\Delta)^+}$, $t\in [0,T]$. In particular the underlying process $S=B^H$ is no longer adapted to the above filtration. 

For the continuous filtration $\mathcal F^W_{(t-\Delta)^+}$, $t\in [0,T]$, consider the corresponding optional projection (see Chapter V in \cite{H}) of $B^H$ 
$$\hat S_t:=\mathbb E\left[B^H_t|\mathcal F^W_{(t-\Delta)^+}\right]=
\int_{0}^{(t-\Delta)^{+}} Z_H(t,s)dW_s, \ \ t\in [0,T].
$$
The Fubini theorem gives that for any process
$\gamma\in L^2(dt\otimes\mathbb P)$ which is progressively measurable with respect to $\mathcal F^W_{(t-\Delta)^+}$, $t\in [0,T]$
we have 
 $
\mathbb E\left[\int_{0}^T \gamma_t B^H_t dt\right]=\mathbb E\left[\int_{0}^T \gamma_t \hat S_t dt\right].
$
Hence, we can apply Theorem \ref{thm2.1} for the optional projection $\hat S$. 

From the Fubini theorem 
$$\int_{0}^T\hat S_t dt=\int_{0}^{T-\Delta}  \left(\int_{s+\Delta}^T Z_H(u,s)du\right)dW_s. $$
Thus, 
the martingale $M$ defined in (\ref{mart}) is equal to 
$$M^{H,\Delta,-}_t=\int_{0}^{(t-\Delta)^{+}}  \left(\int_{s+\Delta}^T Z_H(u,s)du\right)dW_s, \ \ t\in [0,T]$$ 
and so, the optimal strategy is given by 
$$\hat\phi^{H,\Delta,-}_t=\int_{0}^{(t-\Delta)^{+}}  \left(\frac{\int_{s+\Delta}^T Z_H(u,s)du}{T-\Delta-s}-Z_H(t,s)\right)dW_s, \ \ t\in [0,T]. $$
Finally, the corresponding value is given by 
$$
\mathbb E\left[V^{0,\hat\phi^{H,\Delta,-}}_T\right]=
\int_{0}^T \int_{0}^{(t-\Delta)^{+}}  Z^2_H(t,s) ds dt -\int_{0}^{T-\Delta}\frac{\left(\int_{s+\Delta}^T Z_H\left(u,s\right)du\right)^2}{T-\Delta-s} ds.  
$$
\subsection{Insider Information}
Rather than having access to just the natural
augmented filtration $(\mathcal F^W_{t})_{t\geq 0}$ for making decisions the
investor can peek $\Delta \in (0,T]$ time units into the future, and so her information
flow is given by the filtration
$(\mathcal F^W_{t+\Delta})_{t\geq 0}$.

The martingale $M$ defined in (\ref{mart}) is equal to 
$$M^{H,\Delta,+}_t=\int_{0}^{(t+\Delta)\wedge T} \left(\int_{s}^T Z_H(u,s)du\right)dW_s  , \ \ t\in [0,T].$$ 
Hence, 
the optimal strategy is given by 
\begin{eqnarray*}
&\hat\phi^{H,\Delta,+}_t=\frac{1}{T}\int_{0}^{\Delta} \left(\int_{s}^T Z_H(u,s)du\right)dW_s\\
&+\int_{\Delta}^{(t+\Delta)\wedge T} 
\frac{\int_{s}^T Z_H(u,s)du}{T+\Delta-s}dW_s-\int_{0}^t Z_H(t,s) dW_s, \ \ t\in [0,T]
\end{eqnarray*}
and the corresponding value is given by 
\begin{eqnarray*}\label{3.6}
&\mathbb E\left[V^{0,\hat\phi^{H,\Delta,+}}_T\right]=
\int_{0}^T \int_{0}^{t}  Z^2_H(t,s) ds dt -\frac{\left|M^{H,\Delta,+}_0\right|^2}{T}-\int_{\Delta}^{T}\frac{\left(\int_{s}^T Z_H(u,s)du\right)^2}{T+\Delta-s} ds\\
&=\frac{T^{2 H+1}}{2H+1}-\frac{1}{T}\int_{0}^{\Delta} \left(\int_{s}^T Z_H(u,s)du\right)^2ds
-\int_{\Delta}^{T}\frac{\left(\int_{s}^T Z_H(u,s)du\right)^2}{T+\Delta-s} ds.
\end{eqnarray*}

\begin{remark}
Observe that the calculations of this section can be done in a similar way for 
any square integrable Gaussian-Volterra process with RCLL paths and the following property:
The process generates the same filtration as the underlying Brownian motion. This property was studied in details 
in \cite{MISH1,MISH2}.
In this paper we focus on the case where the risky asset is given by a fractional Brownian motion. 
In particular, we apply the obtained formulas 
in order to study numerically the value of the liquidation problem (for different flows of information)
as a function of the Hurst parameter.
\end{remark}

\begin{figure}
    \centering
    \includegraphics[width=0.72\linewidth]{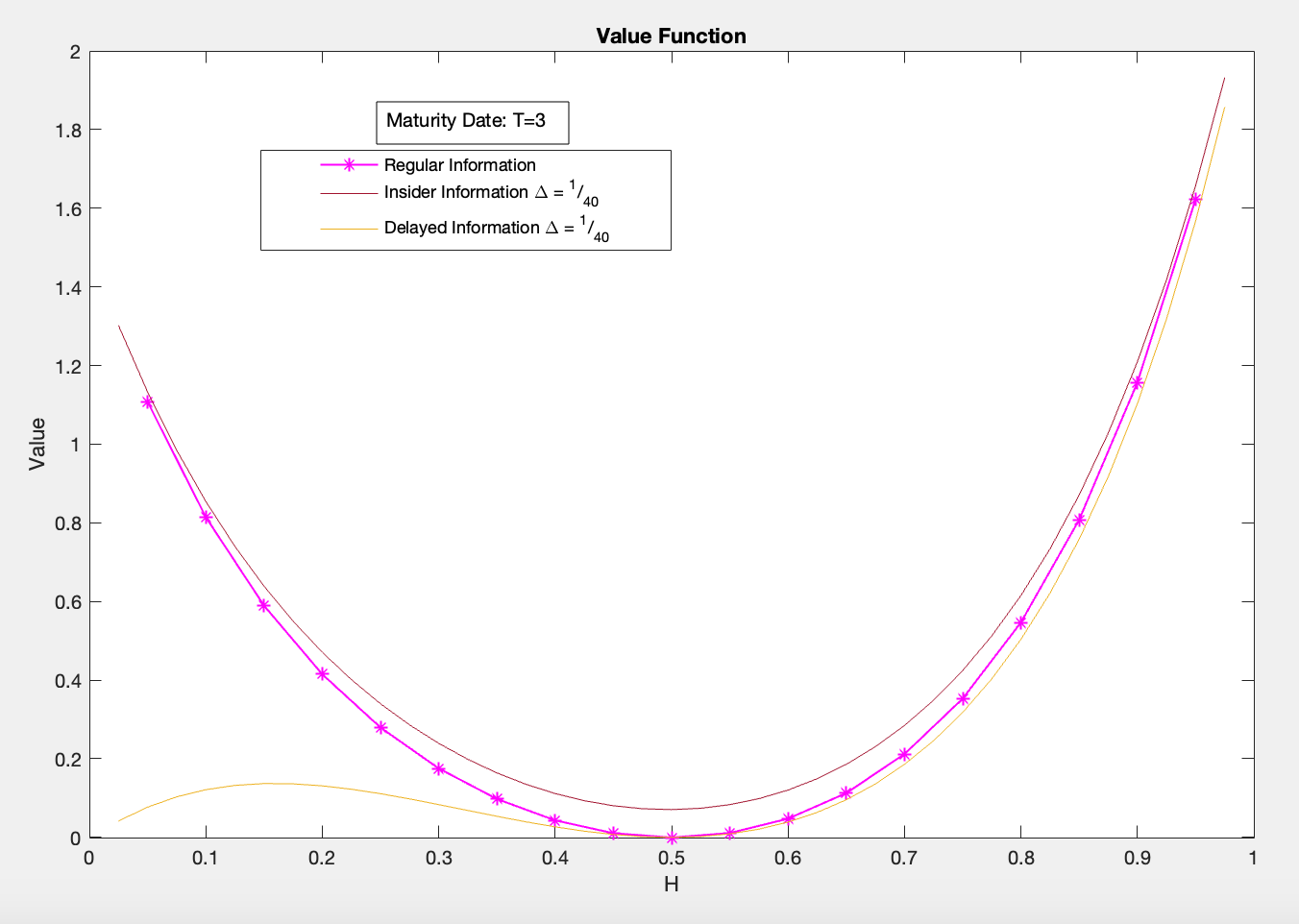}
    \caption{\footnotesize 
    The value of the liquidation problem for different flows of information (shown in different colors)
as a function of the Hurst parameter $H$. Observe that for delayed information the value function is no longer decreasing 
for $H<0.5$.
The reason is that
for very low $H$ values the correlation between the increments decays faster to
$0$ with their time distance, hence a delay results in almost complete loss of
information regarding the current price.}
  \label{fig:enter-label}
\end{figure}

\begin{figure}
    \centering
    \includegraphics[width=0.72\linewidth]{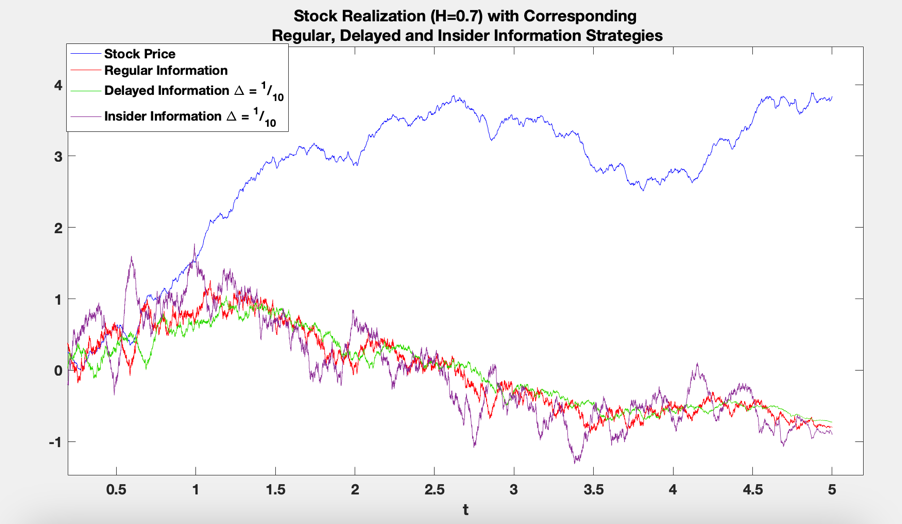}
    \caption{\footnotesize  In this figure we simulate a sample path of a fractional Brownian motion with Hurst parameter $H=0.7$
    and the corresponding optimal trading strategies (we take maturity date $T=5$).
   We observe that
 the Regular Information graph, is a “lagged version” of the Insider Information graph and the Delayed Information graph is a “lagged version” of the Regular Information graph.  
  }
    \label{fig:enter-label}
\end{figure}

\section*{Acknowledgements}
\noindent
Supported in part by the ISF grant 230/21.

\end{document}